\documentclass{article}
\usepackage{amssymb}
\usepackage{amsfonts}
\usepackage{amsmath}
\usepackage[doublespacing]{setspace}

\newtheorem{theorem}{Theorem}
\newtheorem{acknowledgement}[theorem]{Acknowledgement}

\input{tcilatex}

\begin{document}

\title{Position-dependent-mass; Cylindrical coordinates, separability, exact
solvability, and $\mathcal{PT}$-symmetry}
\author{Omar Mustafa \\
Department of Physics, Eastern Mediterranean University, \\
G Magusa, North Cyprus, Mersin 10,Turkey\\
E-mail: omar.mustafa@emu.edu.tr\\
\ Tel: +90 392 630 1314,\\
\ Fax: +90 392 3651604}
\maketitle

\begin{abstract}
The kinetic energy operator with position-dependent-mass in cylindrical
coordinates is obtained. The separability of the corresponding Schr\"{o}%
dinger equation is discussed within radial cylindrical mass settings.
Azimuthal symmetry is assumed and spectral signatures of various $z$%
-dependent interaction potentials (Hermitian and non-Hermitian $\mathcal{PT}$%
-symmetric) are reported.

\medskip PACS codes: 03.65.Ge, 03.65.Ca

Keywords: Position-dependent-mass, cylindrical setting, separability, exact
solvability, $\mathcal{PT}$-symmetry.
\end{abstract}

\section{Introduction}

The von Roos Hamiltonian for position-dependent-mass (PDM) quantum particles
is known to be associated with an ordering ambiguity problem manifested by
the non-unique representation of the kinetic energy operator [1]. In such
Hamiltonian%
\begin{equation}
H=-\frac{\hbar ^{2}}{4}\left[ m\left( \vec{r}\right) ^{\gamma }\vec{\nabla}%
m\left( \vec{r}\right) ^{\beta }\mathbf{\cdot }\vec{\nabla}m\left( \vec{r}%
\right) ^{\alpha }+m\left( \vec{r}\right) ^{\alpha }\;\vec{\nabla}m\left( 
\vec{r}\right) ^{\beta }\mathbf{\cdot }\vec{\nabla}m\left( \vec{r}\right)
^{\gamma }\right] +V\left( \vec{r}\right) ,
\end{equation}%
an obvious profile change in the effective potential is introduced when the
parametric values of the ambiguity parameters $\left( \alpha ,\beta ,\gamma
\right) $\ are changed (within the von Roos constraint $\alpha +\beta
+\gamma =-1$). Nevertheless, it is known that the continuity conditions at
the heterojunction boundaries between two crystals imply $\alpha =\gamma $
(cf., e.g., ref.[2] and the related references cited therein) . This would
effectively reduce the domain of the acceptable parametric values of the
ambiguity parameters. In fact, the PDM Hamiltonian (1) is known to be a
descriptive model for many physical problems (like but not limited to,
many-body problem, electronic properties of semiconductors, etc.) [1-30]. It
is, moreover, a mathematically challenging and a useful model that enriches
the class of exactly solvable quantum mechanical systems.

In the literature, nevertheless, one may find many suggestion on the
ambiguity parametric values. For example, Gora and William have suggested $%
\beta =\gamma =0,$ $\alpha =-1$, Ben Daniel and Duke $\alpha =\gamma =0,$ $%
\beta =-1$, Zhu and Kroemer $\alpha =\gamma =-1/2,$ $\beta =0$, Li and Kuhn $%
\beta =\gamma =-1/2,$ $\alpha =0$, and Mustafa and Mazharimousavi $\alpha
=\gamma =-1/4,$ $\beta =-1/2$ (cf., e.g., [2,3] and references therein).
Very recently, we have studied the problem of a singular PDM particle in an
infinite potential well and shown that none of the above known parametric
ordering sets is admissible within the methodical proposal discussed in [3].
Consequently, the ordering ambiguity conflict does not only depend on the
heterojunction boundaries and the Dutra and Almeida's [4] reliability test
(cf., e.g., Ref.s [3,4] for more details). The potential and/or the form of
the position dependent mass have their say in the process [3]. At the end of
the day, however, the consensus is that this ambiguity is mainly attributed
to the lack of the Galilean invariance (cf., e.g., Ref.[1] on the details of
this issue).

In the current methodical proposal, we shall be working with the ambiguity
parameters as they are without any discrimination as to which set of
ordering is favorable than which. We discuss the von Roos Hamiltonian (1)
using cylindrical coordinates and seek some feasible separability in section
2. Therein, we suggest the position-dependent-mass to be only
radial-dependent (i.e., $m\left( \vec{r}\right) =m_{\circ }M\left( \rho
,\varphi ,z\right) =M\left( \rho ,\varphi ,z\right) =M\left( \rho \right)
=1/\rho ^{2}$) and azimuthal symmetrization is sought through the assumption
that 
\begin{equation}
V\left( \vec{r}\right) =V\left( \rho ,\varphi ,z\right) =\frac{\rho ^{2}}{2}%
\left[ \tilde{V}\left( \rho \right) +\tilde{V}\left( z\right) \right] .
\end{equation}%
Of course, this constitutes only a one feasible separability of the system
(other separability options may occur as well), as justified in section 2.
In section 3, within the radial cylindrical settings,\ we consider two
examples of fundamental nature. The radial cylindrical "Coulombic" $\tilde{V}%
\left( \rho \right) =-2/\rho $ and the "harmonic oscillator" $\tilde{V}%
\left( \rho \right) =a^{2}\rho ^{2}/4$. The spectral signatures of different 
$\tilde{V}\left( z\right) $ settings on the Coulombic and harmonic
oscillator spectra are reported for impenetrable walls at $z=0$ and $z=L$,
for a Morse [31], for a non-Hermitian $\mathcal{PT}$-symmetrized Scarf II
[28,32,33], and for a non-Hermitian $\mathcal{PT}$-symmetrized Samsonov
[28,34] interaction models. Where, $\mathcal{P}$ denotes parity and $%
\mathcal{T}$ mimics the time reflection (cf., e.g., Ref.[28] and references
cited therein on this issue). Our concluding remarks are in section 4.

\section{Cylindrical coordinates and separability}

Let us consider the kinetic energy operator of the PDM Hamiltonian in (1)
and a PDM function of the form $m\left( \vec{r}\right) =m_{\circ }M\left(
\rho ,\varphi ,z\right) =$ $M\left( \rho ,\varphi ,z\right) $ (where $\hbar
=m_{\circ }=1$ units are to be used hereinafter). Moreover, we consider the
following substitutions%
\begin{equation}
\begin{tabular}{l}
$\vec{A}=\alpha M\left( \rho ,\varphi ,z\right) ^{\alpha -1}\left[ \hat{\rho}%
\ \partial _{\rho }+\frac{\hat{\varphi}}{\rho }\partial _{\varphi }+\hat{z}\
\partial _{z}\right] M\left( \rho ,\varphi ,z\right) ,$ \\ 
$\vec{B}=\beta M\left( \rho ,\varphi ,z\right) ^{\beta -1}\left[ \hat{\rho}\
\partial _{\rho }+\frac{\hat{\varphi}}{\rho }\partial _{\varphi }+\hat{z}\
\partial _{z}\right] M\left( \rho ,\varphi ,z\right) ,$ \\ 
$\vec{C}=\gamma M\left( \rho ,\varphi ,z\right) ^{\gamma -1}\left[ \hat{\rho}%
\ \partial _{\rho }+\frac{\hat{\varphi}}{\rho }\partial _{\varphi }+\hat{z}\
\partial _{z}\right] M\left( \rho ,\varphi ,z\right) ,$%
\end{tabular}%
\end{equation}%
to imply%
\begin{equation}
\begin{tabular}{l}
$\vec{\nabla}M\left( \rho ,\varphi ,z\right) ^{\alpha }=\vec{A}+M\left( \rho
,\varphi ,z\right) ^{\alpha }$\ $\vec{\nabla}$ \\ 
$\vec{\nabla}M\left( \rho ,\varphi ,z\right) ^{\beta }=\vec{B}+M\left( \rho
,\varphi ,z\right) ^{\beta }$\ $\vec{\nabla}$ \\ 
$\vec{\nabla}M\left( \rho ,\varphi ,z\right) ^{\gamma }=\vec{C}+M\left( \rho
,\varphi ,z\right) ^{\gamma }$\ $\vec{\nabla}$%
\end{tabular}%
.
\end{equation}%
Using the above identities, one (with $M\left( \rho ,\varphi ,z\right)
\equiv M$ for simplicity of notations) may rewrite%
\begin{eqnarray}
M^{\alpha }\;\vec{\nabla}M^{\beta }\mathbf{\cdot }\vec{\nabla}M^{\gamma }
&=&M^{\alpha }\left( \vec{B}\cdot \vec{C}\right) +M^{\alpha +\gamma }\left( 
\vec{B}\cdot \vec{\nabla}\right)  \notag \\
&&+M^{\alpha +\beta }\left[ 2\vec{C}\cdot \vec{\nabla}+\vec{\nabla}\cdot 
\vec{C}\right] +M^{-1}\vec{\nabla}^{2},
\end{eqnarray}%
and%
\begin{eqnarray}
M^{\gamma }\;\vec{\nabla}M^{\beta }\mathbf{\cdot }\vec{\nabla}M^{\alpha }
&=&M^{\gamma }\left( \vec{B}\cdot \vec{A}\right) +M^{\alpha +\gamma }\left( 
\vec{B}\cdot \vec{\nabla}\right)  \notag \\
&&+M^{\gamma +\beta }\left[ 2\vec{A}\cdot \vec{\nabla}+\vec{\nabla}\cdot 
\vec{A}\right] +M^{-1}\vec{\nabla}^{2}.
\end{eqnarray}%
Let use now consider a class of the mass functions defined as%
\begin{eqnarray}
M\left( \rho ,\varphi ,z\right) &=&g\left( \rho \right) f\left( \varphi
\right) k\left( z\right) \text{ }\Longrightarrow \partial _{\rho }M=M_{\rho
}=g_{\rho }\left( \rho \right) f\left( \varphi \right) k\left( z\right) 
\notag \\
&\Rightarrow &\partial _{\rho }^{2}M=M_{\rho \rho }=g_{\rho \rho }\left(
\rho \right) f\left( \varphi \right) k\left( z\right) .
\end{eqnarray}%
Which would, in effect, imply that the PDM Schr\"{o}dinger equation $\left[
H-E\right] \Psi \left( \rho ,\varphi ,z\right) =0$ for Hamiltonian (1) be
written as%
\begin{eqnarray}
&&\left\{ \partial _{\rho }^{2}+\left( \frac{1}{\rho }-\frac{M_{\rho }}{M}%
\right) \partial _{\rho }+\frac{1}{\rho ^{2}}\left( \partial _{\varphi }^{2}-%
\frac{M_{\varphi }}{M}\partial _{\varphi }\right) +\left( \partial _{z}^{2}-%
\frac{M_{z}}{M}\partial _{z}\right) \right\} \Psi \left( \rho ,\varphi
,z\right)  \notag \\
&=&\left\{ 2MV\left( \rho ,\varphi ,z\right) -2ME-MW\left( \rho ,\varphi
,z\right) \right\} \Psi \left( \rho ,\varphi ,z\right) ,
\end{eqnarray}%
where%
\begin{eqnarray}
2MW\left( \rho ,\varphi ,z\right) &=&\frac{\zeta }{M^{2}}\left[ M_{\rho
}^{2}+\frac{M_{\varphi }^{2}}{\rho ^{2}}+M_{z}^{2}\right]  \notag \\
&&-\frac{\left( \beta +1\right) }{M}\left[ \frac{M_{\rho }}{\rho }+M_{\rho
\rho }+\frac{M_{\varphi \varphi }}{\rho ^{2}}+M_{zz}\right]  \notag \\
&=&\zeta \left[ \left( \frac{g^{\prime }\left( \rho \right) }{g\left( \rho
\right) }\right) ^{2}+\frac{1}{\rho ^{2}}\left( \frac{f^{\prime }\left(
\varphi \right) }{f\left( \varphi \right) }\right) ^{2}+\left( \frac{%
k^{\prime }\left( z\right) }{k\left( z\right) }\right) ^{2}\right]  \notag \\
&&-\left( \beta +1\right) \left[ \frac{g^{\prime }\left( \rho \right) }{\rho
g\left( \rho \right) }+\frac{g^{^{\prime \prime }}\left( \rho \right) }{%
g\left( \rho \right) }+\frac{1}{\rho ^{2}}\frac{f^{^{\prime \prime }}\left(
\varphi \right) }{f\left( \varphi \right) }+\frac{k^{^{\prime \prime
}}\left( z\right) }{k\left( z\right) }\right] .
\end{eqnarray}%
\begin{equation}
\zeta =\alpha \left( \alpha -1\right) +\gamma \left( \gamma -1\right) -\beta
\left( \beta +1\right) .
\end{equation}

At this point, one should notice that the choice of the mass function in (7)
is inspired by the appearance of terms like $\left( M_{\rho }/M\right) $, $%
\left( M_{\varphi }/M\right) $, and $\left( M_{z}/M\right) $ as
multiplicities of the first-order derivatives in (8). This would, in fact,
make the separability of (8) highly feasible and far less complicated.
Moreover, following the traditional general wave function assumption,%
\begin{equation}
\Psi \left( \rho ,\varphi ,z\right) =R\left( \rho \right) \Phi \left(
\varphi \right) Z\left( z\right) ;\text{ }\rho \in \left( 0,\infty \right)
,\ \varphi \in \left( 0,2\pi \right) ,z\in \left( -\infty ,\infty \right)
\end{equation}%
to ease coordinates separability of (8), we obtain%
\begin{eqnarray}
0 &=&2g\left( \rho \right) f\left( \varphi \right) k\left( z\right) \left[
E-V\left( \rho ,\varphi ,z\right) \right]  \notag \\
&&+\left[ \frac{R^{\prime \prime }\left( \rho \right) }{R\left( \rho \right) 
}-\left( \frac{g^{\prime }\left( \rho \right) }{g\left( \rho \right) }-\frac{%
1}{\rho }\right) \frac{R^{\prime }\left( \rho \right) }{R\left( \rho \right) 
}\right.  \notag \\
&&+\left. \frac{\zeta }{2}\left( \frac{g^{\prime }\left( \rho \right) }{%
g\left( \rho \right) }\right) ^{2}-\frac{\left( \beta +1\right) }{2}\left( 
\frac{g^{\prime }\left( \rho \right) }{\rho g\left( \rho \right) }+\frac{%
g^{^{\prime \prime }}\left( \rho \right) }{g\left( \rho \right) }\right) %
\right]  \notag \\
&&+\left[ \frac{Z^{\prime \prime }\left( z\right) }{Z\left( z\right) }-\frac{%
k^{\prime }\left( z\right) }{k\left( z\right) }\frac{Z^{\prime }\left(
z\right) }{Z\left( z\right) }+\frac{\zeta }{2}\left( \frac{k^{\prime }\left(
z\right) }{k\left( z\right) }\right) ^{2}-\frac{\left( \beta +1\right) }{2}%
\frac{k^{^{\prime \prime }}\left( z\right) }{k\left( z\right) }\right] 
\notag \\
&&+\frac{1}{\rho ^{2}}\left[ \frac{\Phi ^{\prime \prime }\left( \varphi
\right) }{\Phi \left( \varphi \right) }-\frac{f^{\prime }\left( \varphi
\right) }{f\left( \varphi \right) }\frac{\Phi ^{\prime }\left( \varphi
\right) }{\Phi \left( \varphi \right) }+\frac{\zeta }{2}\left( \frac{%
f^{\prime }\left( \varphi \right) }{f\left( \varphi \right) }\right) ^{2}-%
\frac{\left( \beta +1\right) }{2}\frac{f^{^{\prime \prime }}\left( \varphi
\right) }{f\left( \varphi \right) }\right]
\end{eqnarray}

It is obvious that separability is granted through a variety of choices. The
simplest of which may be sought in an obviously "manifested-by-equation
(12)" general identity of the form%
\begin{equation}
2MV\left( \rho ,\varphi ,z\right) =2g\left( \rho \right) f\left( \varphi
\right) k\left( z\right) V\left( \rho ,\varphi ,z\right) =\tilde{V}\left(
\rho \right) +\tilde{V}\left( z\right) +\frac{1}{\rho ^{2}}\tilde{V}\left(
\varphi \right) .
\end{equation}%
In this case, we may avoid any specifications on the forms of $g\left( \rho
\right) $, $f\left( \varphi \right) $, and $k\left( z\right) $ rather than
being mathematically and quantum mechanically "very well" defined. However,
the energy term, $2g\left( \rho \right) f\left( \varphi \right) k\left(
z\right) E$, in (12) suggests three feasible separabilities for $f\left(
\varphi \right) =1=k\left( z\right) $, $k\left( z\right) =1=g\left( \rho
\right) $, and $f\left( \varphi \right) =1=g\left( \rho \right) $. We focus
on one of these cases in the sequel.

Let us consider the position-dependent-mass function to be only an explicit
function of $\rho $. Namely, we choose $f\left( \varphi \right) =1=k\left(
z\right) $ and $g\left( \rho \right) =\rho ^{-2}$ so that $M\left( \rho
,\varphi ,z\right) =M\left( \rho \right) =\rho ^{-2}$. Under these settings,
equation(12) collapses into a simple separable form%
\begin{eqnarray}
0 &=&\left[ \frac{\Phi ^{\prime \prime }\left( \varphi \right) }{\Phi \left(
\varphi \right) }+2E-\tilde{V}\left( \varphi \right) +2\left( \zeta -\beta
-1\right) \right]  \notag \\
&&+\rho ^{2}\left[ \frac{R^{\prime \prime }\left( \rho \right) }{R\left(
\rho \right) }+\frac{3}{\rho }\frac{R^{\prime }\left( \rho \right) }{R\left(
\rho \right) }-\tilde{V}\left( \rho \right) +\frac{Z^{\prime \prime }\left(
z\right) }{Z\left( z\right) }-\tilde{V}\left( z\right) \right] .
\end{eqnarray}%
Equation (14) with azimuthal symmetry (i.e., $\tilde{V}\left( \varphi
\right) =0$) would immediately imply that%
\begin{equation}
\left[ \frac{\Phi ^{\prime \prime }\left( \varphi \right) }{\Phi \left(
\varphi \right) }+2E+2\left( \zeta -\beta -1\right) \right] =K_{\varphi
}^{2},
\end{equation}%
and%
\begin{equation}
\left[ \frac{R^{\prime \prime }\left( \rho \right) }{R\left( \rho \right) }+%
\frac{3}{\rho }\frac{R^{\prime }\left( \rho \right) }{R\left( \rho \right) }+%
\frac{K_{\varphi }^{2}}{\rho ^{2}}-\tilde{V}\left( \rho \right) \right] +%
\left[ \frac{Z^{\prime \prime }\left( z\right) }{Z\left( z\right) }-\tilde{V}%
\left( z\right) \right] =0.
\end{equation}%
In due course, the solution of (15) reads $\Phi \left( \varphi \right) =\exp
\left( im\varphi \right) $ where $m=0,\pm 1,\pm 2,\cdots $ is the magnetic
quantum number and $\Phi \left( \varphi \right) $ satisfies the single
valued condition $\Phi \left( \varphi \right) =\Phi \left( \varphi +2\pi
\right) $. Moreover, we obtain%
\begin{equation}
K_{\varphi }^{2}=2E+2\left( \zeta -\beta -1\right) -m^{2}.
\end{equation}%
Consequently, one may cast%
\begin{equation}
\frac{Z^{\prime \prime }\left( z\right) }{Z\left( z\right) }-\tilde{V}\left(
z\right) =-K_{z}^{2}
\end{equation}%
and%
\begin{equation}
\frac{R^{\prime \prime }\left( \rho \right) }{R\left( \rho \right) }+\frac{3%
}{\rho }\frac{R^{\prime }\left( \rho \right) }{R\left( \rho \right) }+\frac{%
K_{\varphi }^{2}}{\rho ^{2}}-\tilde{V}\left( \rho \right) =K_{z}^{2}
\end{equation}%
In the following section, we consider $\tilde{V}\left( \rho \right) $ to
represent a "Coulombic" and a "harmonic oscillator" and find the spectral
signatures of different $\tilde{V}\left( z\right) $ potentials of (18) on
the over all spectra.

\section{Two examples; the radial cylindrical Coulombic and the
harmonic-oscillator}

A priori, we remove the first-order derivative in the radial cylindrical
part of (19) and redefine%
\begin{equation}
R\left( \rho \right) =\rho ^{-3/2}U\left( \rho \right) ,
\end{equation}%
to obtain%
\begin{equation}
-U^{\prime \prime }\left( \rho \right) +\left[ \frac{3/4-K_{\varphi }^{2}}{%
\rho ^{2}}+\tilde{V}\left( \rho \right) \right] U\left( \rho \right)
=-K_{z}^{2}U\left( \rho \right) .
\end{equation}%
In fact, this 1D radial cylindrical Schr\"{o}dinger equation provides an
effective tool to study the effect of different $\tilde{V}\left( z\right) $
settings of (18) on the spectra of two interesting models of fundamental
nature. The Coulombic and the harmonic oscillator [30]. Of course, such
effects could be tested for other models.

Let us take a Coulombic radial cylindrical model $\tilde{V}\left( \rho
\right) =-2/\rho $. In this case, equation (21) would read%
\begin{equation}
-U^{\prime \prime }\left( \rho \right) +\left[ \frac{\ell ^{2}-1/4}{\rho ^{2}%
}-\frac{2}{\rho }\right] U\left( \rho \right) =-K_{z}^{2}U\left( \rho
\right) ,
\end{equation}%
where $\ell =\left( 1-K_{\varphi }^{2}\right) ^{1/2}$, $K_{z}=\left( n_{\rho
}+\ell +1\right) ^{-1}$, and $n_{\rho }=0,1,2,\cdots $ is the radial quantum
number. Hence, $K_{z}=1/\left( n_{\rho }+\sqrt{1-K_{\varphi }^{2}}+1\right) $
and%
\begin{equation}
E=\left( \frac{m^{2}+3}{2}\right) -\left( \zeta -\beta \right) -\frac{1}{2}%
\left( \frac{1}{K_{z}}-n_{\rho }-1\right) ^{2},
\end{equation}%
where $K_{z}$ is to be determined through the solution of (18) under
different $\tilde{V}\left( z\right) $ settings.

Next, we consider the radial cylindrical harmonic oscillator model $\tilde{V}%
\left( \rho \right) =a^{2}\rho ^{2}/4$ in (21) to obtain%
\begin{equation}
E=\left( \frac{m^{2}+3}{2}\right) -\left( \zeta -\beta \right) -\frac{1}{2}%
\left( \frac{K_{z}^{2}}{a}+2n_{\rho }+1\right) ^{2},
\end{equation}%
where, again, $K_{z}$ is to be determined through the solution of (18) under
different $\tilde{V}\left( z\right) $ settings in the sequel subsections.
Nevertheless, it is obvious that the position-dependent-mass spectral
signature is documented through the ambiguity parameters appearance in the
constant shift, (i.e., $\left[ -\left( \zeta -\beta -1\right) \right] $ as
obvious from (15) and (17) and included in the energy eigenvalues of (23)
and (24)).

\subsection{Spectral signature of impenetrable walls at $z=0$ and $z=L$}

Lets us now consider that the above mentioned position-dependent-mass
particle is trapped to move between two impenetrable walls at $z=0$ and $z=L$%
. We may then take%
\begin{equation}
\tilde{V}\left( z\right) =\left\{ 
\begin{tabular}{ll}
$0$ & ; $0<z<L$ \\ 
$\infty $ & ; elsewhere%
\end{tabular}%
\right. .
\end{equation}%
Consequently, equation (18) reads%
\begin{equation}
Z^{\prime \prime }\left( z\right) +K_{z}^{2}Z\left( z\right) =0,
\end{equation}%
where $Z\left( z\right) $ satisfies the boundary conditions $Z\left(
z=0\right) =0=Z\left( z=L\right) $ and implies that%
\begin{equation}
Z\left( z\right) =\sin K_{z}z\ ;\text{ }K_{z}=\frac{n_{z}\pi }{L}\ ,\text{ }%
n_{z}=1,2,3,\cdots .
\end{equation}%
Hence, $K_{z}^{2}=n_{z}^{2}\pi ^{2}/L^{2}$\ and the quantum PDM particle
here is quasi-free in the $z$-direction (i.e., $\tilde{V}\left( z\right) =0$%
) but constrained to move between the two impenetrable walls at $z=0$ and $%
z=L$. The spectral signature of such $z$-dependent potential settings is
clear, therefore. That is, a quantum particle endowed with a
position-dependent-mass $M\left( \rho ,\varphi ,z\right) =M\left( \rho
\right) =\rho ^{-2}$ and subjected to an interaction potential of the form%
\begin{equation}
V\left( \rho ,\varphi ,z\right) =-2\rho +\rho ^{2}\tilde{V}\left( z\right) ,
\end{equation}%
with $\tilde{V}\left( z\right) $ defined in (25), would admit exact energy
eigenvalues given by%
\begin{equation}
E_{n_{\rho },m,n_{z}}=\left( \frac{m^{2}+3}{2}\right) -\left( \zeta -\beta
\right) -\frac{1}{2}\left( \frac{L}{n_{z}\pi }-n_{\rho }-1\right) ^{2},
\end{equation}

On the other hand, a quantum particle endowed with a position-dependent-mass 
$M\left( \rho ,\varphi ,z\right) =M\left( \rho \right) =\rho ^{-2}$
subjected to an interaction potential of the form%
\begin{equation}
V\left( \rho ,\varphi ,z\right) =a^{2}\rho ^{4}/4+\rho ^{2}\tilde{V}\left(
z\right) ,
\end{equation}%
with $\tilde{V}\left( z\right) $ defined in (25), would be accompanied by
exact energy eigenvalues of the form%
\begin{equation}
E_{n_{\rho },m,n_{z}}=\left( \frac{m^{2}+3}{2}\right) -\left( \zeta -\beta
\right) -\frac{1}{2}\left( \frac{n_{z}^{2}\pi ^{2}}{aL^{2}}+2n_{\rho
}+1\right) ^{2}.
\end{equation}

\subsection{Spectral signatures of a $\tilde{V}\left( z\right) $ Morse model}

Consider a Morse type interaction $\tilde{V}\left( z\right) =D\ \left(
e^{-2\epsilon z}-2e^{-\epsilon z}\right) ;$ $D>0$, in (18). We may then
closely follow the methodical proposal of Chen [31] to obtain%
\begin{equation}
K_{z}^{2}=\left( \frac{\sqrt{D}}{\epsilon }-\tilde{n}_{z}-\frac{1}{2}\right)
,\text{ }\tilde{n}_{z}=0,1,2,3,\cdots
\end{equation}%
where one should consider $2m=\hbar =1$, $a\rightarrow \epsilon $, $%
E\rightarrow K_{z}^{2}$ and $x\rightarrow z$ of Chen [31] to match our
settings in (18). Therefore, a PDM quantum particle endowed with $M\left(
\rho ,\varphi ,z\right) =M\left( \rho \right) =\rho ^{-2}$ and subjected to
an interaction potential of the form%
\begin{equation}
V\left( \rho ,\varphi ,z\right) =-2\rho +D\rho ^{2}\ \left( e^{-2\epsilon
z}-2e^{-\epsilon z}\right) ;D>0,
\end{equation}%
would admit exact energy eigenvalues given by%
\begin{equation}
E_{n_{\rho },m,\tilde{n}_{z}}=\left( \frac{m^{2}+3}{2}\right) -\left( \zeta
-\beta \right) -\frac{1}{2}\left( \frac{1}{\sqrt{\frac{\sqrt{D}}{\epsilon }-%
\tilde{n}_{z}-\frac{1}{2}}}-n_{\rho }-1\right) ^{2}.
\end{equation}%
Obviously, the condition $\left( \sqrt{D}/\epsilon -\tilde{n}_{z}-\frac{1}{2}%
\right) >0$ is manifested here and ought to be enforced, otherwise complex
pairs of energy eigenvalues are obtained in the process.

Moreover, a quantum particle with $M\left( \rho ,\varphi ,z\right) =M\left(
\rho \right) =\rho ^{-2}$ subjected to an interaction potential%
\begin{equation}
V\left( \rho ,\varphi ,z\right) =a^{2}\rho ^{4}/4+D\rho ^{2}\ \left(
e^{-2\epsilon z}-2e^{-\epsilon z}\right) ;D>0,
\end{equation}%
would indulge the exact energy eigenvalues%
\begin{equation}
E_{n_{\rho },m,\tilde{n}_{z}}=\left( \frac{m^{2}+3}{2}\right) -\left( \zeta
-\beta \right) -\frac{1}{2}\left( \frac{1}{a}\left[ \frac{\sqrt{D}}{\epsilon 
}-\tilde{n}_{z}-\frac{1}{2}\right] +2n_{\rho }+1\right) ^{2}.
\end{equation}

\subsection{$\mathcal{PT}$-symmetrized $\tilde{V}\left( z\right) $ spectral
signatures}

We may now consider a $\mathcal{PT}$-symmetrized $\tilde{V}\left( z\right) $
Scarf II in (18) so that%
\begin{equation}
\tilde{V}\left( z\right) =-\frac{3+A^{2}}{4\cosh ^{2}z}-i\frac{A\sinh z}{%
\cosh ^{2}z},
\end{equation}%
where the corresponding Hamiltonian is known to be a non-Hermitian $\mathcal{%
PT}$-symmetric Hamiltonian that admits exact eigenvalues (cf., e.g., Mustafa
and Mazharimousavi \{28], Ahmed [32], and Khare [33]) of the form%
\begin{equation}
K_{z}^{2}=\left\{ 
\begin{tabular}{ll}
$-\left( n_{z}+\frac{1-A}{2}\right) ^{2};$ $\text{ }n_{z}=0,1,2,3,\cdots <%
\frac{A-1}{2},$ & for $A\geq 2,$ \\ 
$-\frac{1}{4};$ & for $A<2,$%
\end{tabular}%
\right. .
\end{equation}%
Hence, a quantum particle with $M\left( \rho ,\varphi ,z\right) =M\left(
\rho \right) =\rho ^{-2}$ moving in%
\begin{equation}
V\left( \rho ,\varphi ,z\right) =-2\rho -\rho ^{2}\ \left[ \frac{3+A^{2}}{%
4\cosh ^{2}z}+i\frac{A\sinh z}{\cosh ^{2}z}\right] ,
\end{equation}%
would encounter complex pairs of energy eigenvalues since $K_{z}=i\left(
n_{z}+\frac{1-A}{2}\right) $ (i.e., $E_{n_{\rho },m,n_{z}}\in 
\mathbb{C}
$). Whereas, when the same PDM-particle is moving in%
\begin{equation}
V\left( \rho ,\varphi ,z\right) =a^{2}\rho ^{4}/4-\rho ^{2}\ \left[ \frac{%
3+A^{2}}{4\cosh ^{2}z}+i\frac{A\sinh z}{\cosh ^{2}z}\right] ,
\end{equation}%
it would admit exact and real energy eigenvalues as%
\begin{equation}
E=\left( \frac{m^{2}+3}{2}\right) -\left( \zeta -\beta \right) -\frac{1}{2}%
\left( \frac{K_{z}^{2}}{a}+2n_{\rho }+1\right) ^{2},
\end{equation}%
with $K_{z}^{2}$ defined in (38). Of course, this should never be attributed
to $\mathcal{PT}$-symmetricity or non-$\mathcal{PT}$-symmetricity of the
original Hamiltonian (1) with the attendant complex non-Hermitian settings.
It is very much related to the nature of separability we followed in this
methodical proposal.

One may wish to consider the $\mathcal{PT}$-symmetric Samsonov [34,28] model%
\begin{equation}
\tilde{V}\left( z\right) =-\frac{1}{\cos z+2i\sin z};\ z\in \left[ -\pi ,\pi %
\right] ,
\end{equation}%
in (18). In this case $Z\left( -\pi \right) =Z\left( \pi \right) =0$ and 
\begin{equation}
K_{z}^{2}=n_{z}^{2}/4;n_{z}=1,3,4,\cdots ,
\end{equation}%
with a missing state $n_{z}=2$ (the reader may refer to Samsonov [34] on
more details on this missing state). Hence, for a PDM quantum particle
endowed with $M\left( \rho ,\varphi ,z\right) =M\left( \rho \right) =\rho
^{-2}$ and subjected to an interaction potential of the form%
\begin{equation}
V\left( \rho ,\varphi ,z\right) =-2\rho -\frac{1}{\cos z+2i\sin z};\ z\in %
\left[ -\pi ,\pi \right] ,
\end{equation}%
the exact energy eigenvalues would read%
\begin{equation}
E_{n_{\rho },m,n_{z}}=\left( \frac{m^{2}+3}{2}\right) -\left( \zeta -\beta
\right) -\frac{1}{2}\left( \frac{2}{n_{z}}-n_{\rho }-1\right) ^{2}.
\end{equation}%
Whereas, for%
\begin{equation}
V\left( \rho ,\varphi ,z\right) =a^{2}\rho ^{4}/4-\frac{1}{\cos z+2i\sin z}%
;\ z\in \left[ -\pi ,\pi \right] ,
\end{equation}%
the exact energy eigenvalues would read%
\begin{equation}
E_{n_{\rho },m,n_{z}}=\left( \frac{m^{2}+3}{2}\right) -\left( \zeta -\beta
\right) -\frac{1}{2}\left( \frac{n_{z}^{2}}{4a}+2n_{\rho }+1\right) ^{2}.
\end{equation}%
where $\tilde{n}_{z}=1,3,4,\cdots $.

\section{Concluding remarks}

The kinetic energy operator in the PDM Hamiltonian (1) is a problem with
many aspects that are yet to be explored. In the current work, we tried to
study this problem within the context of cylindrical coordinates $\left(
\rho ,\varphi ,z\right) $. In due course, the essentials related with the
kinetic energy operator in (1) are reported. The separability of the Schr%
\"{o}dinger equation is sought through a radial cylindrical position
dependent mass $M\left( \rho ,\varphi ,z\right) =M\left( \rho \right)
=1/\rho ^{2}$ accompanied by an azimuthally symmetrized interaction
potential $V\left( \rho ,\varphi ,z\right) =\rho ^{2}\left[ \tilde{V}\left(
\rho \right) +\tilde{V}\left( z\right) \right] /2$, where $\tilde{V}\left(
\varphi \right) =0$. Such a combination is not a unique one and some other
separability settings could be sought. However, we have chosen to stick with
the above mentioned combination for it leads into a handy though rather
constructive separable system of the one-dimensional Schr\"{o}dinger
equations (15), (18), and (19).

Assuming azimuthal symmetrization of the problem at hand and within the
radial settings,\ we consider two examples of fundamental nature. The radial
cylindrical Coulombic $\tilde{V}\left( \rho \right) =-2/\rho $ and the
radial cylindrical harmonic oscillator $\tilde{V}\left( \rho \right)
=a^{2}\rho ^{2}/4$. They are indeed exactly solvable within the settings of
(21) and admit exact energy eigenvalues documented in (23) and (24),
respectively. Nevertheless, the appearance of $K_{z}$ and $K_{z}^{2}$ in
(23) and (24), respectively, offered an opportunity to study their spectral
signatures mandated by different $\tilde{V}\left( z\right) $ interaction
models. Namely, the spectral signatures of $\tilde{V}\left( z\right) $ for
impenetrable walls at $z=0$ and $z=L$ (27), for a Morse (32), for a
non-Hermitian $\mathcal{PT}$-symmetrized Scarf II (38), and for a
non-Hermitian $\mathcal{PT}$-symmetrized Samsonov [28,34] (43) are reported.

To summarize, we have assumed azimuthal symmetry and used the radial
cylindrical Coulomb and harmonic oscillator to obtain exact eigenvalues for
a new set of interaction potentials (represented in their general form in
(2) and detailed in (28), (30), (33), (35), (39), (40), (44), and (46)). In
fact, under such azimuthal symmetrization and $\tilde{V}\left( z\right) $
setting, this set of exactly-solvable models may grow up as long as one can
find exactly-solvable radially cylindrical models (hereby, exact-solvability
may even include numerically exactly-solvable models as well). The recipe as
how to collect the energy eigenvalues is clear in the above methodical
proposal.

\begin{acknowledgement}
I would like to thank both referees for their valuable comments and
suggestions.\newpage
\end{acknowledgement}


\begin{thebibliography}{99}
\bibitem{} O von Roos, Phys. Rev. \textbf{B 27} (1983) 7547.

\bibitem{} O Mustafa, S.Habib Mazharimousavi, Int. J. Theor. Phys \ \textbf{%
46} (2007) 1786.

R. Koc, G. Sahinoglu, M. Koca, Eur. Phys. J. \textbf{B 48} (2005) 583.

\bibitem{} O Mustafa, S H Mazharimousavi, Phys. Lett. \textbf{A 373} (2009)
325.

\bibitem{} A de Souza Dutra, C A S Almeida, Phys Lett. \textbf{A 275} (2000)
2

\bibitem{} A Puente, M Casas, Comput. Mater Sci. \textbf{2} (1994) 441

\bibitem{} A R Plastino, M Casas, A Plastino, Phys. Lett. \textbf{A281}
(2001) 297.

\bibitem{} A Schmidt, Phys. Lett. \textbf{A 353} (2006) 459.

A Schmidt, J Phys \textbf{A: }Math. Theor.\textbf{42 (}2009\textbf{) }245304.

\bibitem{} S H Dong, M Lozada-Cassou, Phys. Lett. \textbf{A 337} (2005) 313.

\bibitem{} I O Vakarchuk, J. Phys. \textbf{A}; Math. Gen. \textbf{38} (2005)
4727.

\bibitem{} C Y Cai, Z Z Ren, G X Ju, Commun. Theor. Phys. \textbf{43} (2005)
1019.

\bibitem{} B Roy, P Roy, Phys. Lett. \textbf{A 340} (2005) 70.

\bibitem{} B Gonul, M Kocak, Chin. Phys. Lett. \textbf{20} (2005) 2742.

\bibitem{} S. Cruz y Cruz, J Negro, L. M. Nieto, Phys. Lett. \textbf{A 369}
(2007) 400.

\bibitem{} S. Cruz y Cruz, O Rosas-Ortiz, J Phys \textbf{A}: Math. Theor. 
\textbf{42} (2009) 185205

\bibitem{} J Lekner, Am. J. Phys. \textbf{75} (2007) 1151

\bibitem{} C Quesne, V M Tkachuk, J. \ Phys. \textbf{A}: Math. Gen. \textbf{%
37} (2004) 4267.

\bibitem{} L Jiang, L Z Yi, C S Jia, Phys. Lett. \textbf{A 345} (2005) 279.

\bibitem{} O Mustafa, S H Mazharimousavi, Phys. Lett. \textbf{A 358} (2006)
259.

\bibitem{} J I Diaz, J Negro, L M Nieto, O Rosas-Ortiz, J Phys \textbf{A};
Math. Gen. \textbf{32} (1999) 8447

\bibitem{} A D Alhaidari, Phys. Rev. \textbf{A 66} (2002) 042116.

\bibitem{} O Mustafa, S H Mazharimousavi, J. Phys. \textbf{A}: Math. Gen. 
\textbf{39} (2006) 10537.

\bibitem{} B Bagchi, A Banerjee, C Quesne, V M Tkachuk, J. \ Phys. \textbf{A}%
; Math. Gen. \textbf{38} (2005) 2929.

\bibitem{} J Yu, S H Dong, Phys. Lett. \textbf{A 325} (2004) 194.

\bibitem{} C Quesne, Ann. Phys. \textbf{321} (2006) 1221.

\bibitem{} T Tanaka, J. Phys. \textbf{A}; Math. Gen. \textbf{39} (2006) 219.

\bibitem{} A de Souza Dutra, J. Phys. \textbf{A}; Math. Gen. \textbf{39}
(2006) 203.

A de Souza Dutra, J A Oliveira, J Phys \textbf{A: }Math. Theor.\textbf{42 (}%
2009\textbf{) }025304.

\bibitem{} O Mustafa, S H Mazharimousavi, Czech. J. Phys \textbf{56} (2006)
297.

\bibitem{} O Mustafa, S H Mazharimousavi, Phys. Lett. \textbf{A 357} (2006)
295.

\bibitem{} O Mustafa, S H Mazharimousavi, J Phys \textbf{A: }Math. Theor.%
\textbf{41 (}2008\textbf{) }244020.

\bibitem{} O Mustafa, J Phys.: Condens. Matter \textbf{5} (1993) 1327.

\bibitem{} G. Chen, Phys. Lett. \textbf{A 326} (2004) 55.

\bibitem{} Z. Ahmed, Phys. Lett. \textbf{A 282} (2001) 343.

\bibitem{} A. Khare, Phys. Lett. \textbf{A 288} (2001) 69.

\bibitem{} B. F. Samsonov, P. Roy, J. Phys. \textbf{A}; Math. Gen. \textbf{38%
} (2005) L249.
\end{thebibliography}
\end{document}